\documentclass[twocolumn]{article}
\usepackage[margin=2cm]{geometry} 
\usepackage{authblk} 
\usepackage{graphicx} 
\usepackage{biblatex} 
\usepackage{subcaption}
\usepackage{physics} 
\usepackage{xcolor} 
\usepackage{hyperref}
\usepackage{amssymb}
\usepackage{threeparttable}

\title{Gravitational Waves from Magnetically Confined Mountains on Accreting White Dwarfs}
\author[1]{Pedro H. B. Rossetto}
\author[2]{Manoel F. Sousa}
\author[1]{Diego A. Falceta-Gonçalves}
\affil[1]{Escola de Artes, Ciências e Humanidades, Universidade de São Paulo \\ Rua Arlindo Bettio, CEP 03828-000, 1000 São Paulo, SP, Brazil}
\affil[2]{Instituto de Física de São Carlos, Universidade de São Paulo, Av. Trabalhador São-carlense 400, São Carlos, SP, Brazil}
\date{}

\begin{document}
\twocolumn[{
\maketitle
\begin{abstract}
    We propose fast-spinning, magnetic white dwarfs in binary systems as promising sources of continuous gravitational waves for future space-based detectors. The aim of this work is to determine the maximum ellipticities and gravitational wave amplitudes achievable by magnetically confined mountains accreted on the star's magnetic poles, and to assess their detectability against predicted sensitivity curves. Self-consistent magnetohydrodynamic equilibrium simulations are used to characterize the structure of the accreted mountain. A parametric search is then performed on these simulations to determine the maximum mass values that can be confined by the star's magnetic field. The stellar ellipticity $\varepsilon$ grows nearly linearly with accreted mass $M_{\rm a}$ up to a maximum, and was found to scale as
$\varepsilon_{\rm max} \simeq 1.6\times10^{-2}T_5^{-2} B_{*,8}^{2}$, while the maximum gravitational wave amplitude as 
$h_{\rm 0, max} \simeq 1.2\times10^{-22}T_5^{-2} B_{*,8}^{2}$, where $T_5$ represents the temperature of the mountain in units of $10^5 \mathrm{K}$ and $B_{*,8}$ represents the star's polar magnetic field intensity in units of $10^8 \mathrm{G}$.
Across a representative range of source inclinations, distances and spin periods, magnetically confined mountains on white dwarfs can generate continuous gravitational waves detectable by future space-based interferometers. BBO and DECIGO can probe diverse configurations, whereas LISA, TianQin, and Taiji require highly magnetized systems. Applications to known fast-spinning white dwarfs indicate that cooler accreted mountains favor detectability.

\,

\end{abstract}
\vspace{2em} 
}]

\section{Introduction}
Rotating compact stars with non-axisymmetric mass distributions are potential sources of continuous gravitational waves (CWs), as the mass quadrupole is not stationary in the inertial frame. Neutron stars (NSs) are expected to present such asymmetries due to crustal deformations, internal magnetic stresses, rotational instabilities, or the accumulation of matter from accretion \cite{ushomirsky2000deformations,gittins2020modelling,gourgoulhon1996gravitational,payne2004burial,haskell2015detecting,riles2023searches}. Among these mechanisms, magnetically confined mountains are particularly relevant - possibly dominant -, as accreted material is channeled by the magnetic field towards the stellar magnetic poles. The resulting pressure gradients distort the magnetic flux surfaces and redistribute the accreted matter over the stellar surface. As accretion proceeds, the magnetic field can be compressed and partially buried. This process, commonly referred to as magnetic burial, establishes a coupling between the accumulated mass, the magnetic-field configuration, and the gravitational-wave-emitting deformation \cite{payne2004burial,melatos2005gravitational,vigelius2009continuous,rossetto2023magnetically}.

Studies of mountains in NSs have shown that the ellipticity is not determined solely by the total accreted mass, but also by the spatial distribution of matter and by the magnetic stresses that confine it. These results emphasize that the calculation of a gravitational-wave amplitude requires a self-consistent determination of the stellar equilibrium and of the associated mass moments, rather than an assumed polar mass distribution (representing the mountain). The distinction is particularly important when assessing whether a computed deformation corresponds to a persistent, radiatively relevant non-axisymmetric structure \cite{priymak2011quadrupole,vigelius2009improved,gittins2024gravitational}.

White dwarfs (WDs) represent a complementary --- and possibly more accessible --- scenario to investigate the physics of CWs generated by magnetic mountains. Like neutron stars, these objects can occur in binary systems, where accretion of matter, together with strong magnetic fields and fast rotation, complete the recipe for a non-axisymmetric quadrupole to produce CWs. The possibility of gravitational-wave emission from rapidly rotating and magnetized WDs has been investigated using both accretion-induced asymmetries and magnetic deformation. In particular, Sousa et al. (2024) considered rapidly rotating magnetic WDs and found that the contribution of magnetic deformation to the predicted CW amplitude is generally negligible when compared with the contribution associated with accreted matter. Their results suggest that magnetic deformation alone is unlikely to be the dominant emission mechanism in the systems considered. Nevertheless, in their simplified modeling, the non-axisymmetric mass distribution due to accretion was prescribed based on the accumulation of matter near the magnetic poles, but does not resolve the full magnetohydrodynamic equilibrium and redistribution of the accreted material  \cite{sousa2020gravitational,sousa2024prospects}. This motivates a more detailed investigation of magnetically confined mountains in WDs, in which the magnetic configuration, density distribution, mass quadrupole, and magnetic burial are calculated within a common equilibrium framework.

In this work, we extend the modeling of \citeauthor{rossetto2026magnetically} (\citeyear{rossetto2026magnetically}) \cite{rossetto2026magnetically}, where we have presented axisymmetric magnetohydrodynamic equilibria for magnetically confined mountains on white dwarfs, to investigate their implications for the stellar mass ellipticity and gravitational-wave emission. As prescribed by \cite{rossetto2026magnetically}, the magnetic-field configuration and the distribution of accreted matter are obtained from a Grad–Shafranov formulation, allowing deformation and magnetic-field rearrangement to be determined self-consistently. In this work, however, we focus on the relationship between the accreted mass, the magnetic-field geometry, and the resulting ellipticity. Particular attention is given to the distinction between an axisymmetric deformation about the magnetic axis and the non-axisymmetric component required to generate a time-dependent quadrupole in the inertial frame. By extending the magnetically confined mountain framework to white dwarfs, this study aims to clarify whether accretion-driven magnetic confinement can generate a physically significant quadrupole and to establish the conditions under which such a deformation could contribute to continuous gravitational-wave emission.

The manuscript is organized as follows. In section \ref{sec:mountains_theory}, we present the theory of magnetically confined mountains due to accretion onto WDs, where we derive the basic equations and describe the Grad-Shafranov method for steady-state solutions. In Section \ref{sec:gw_model}, we describe the ellipticity of WDs with magnetically confined mountains, and the model of gravitational wave emission from the quadrupole mass distribution of such asymmetric WDs. The main results of this work are presented in Section \ref{sec:results}, followed by discussions and conclusions in Section \ref{sec:conclusions}.

\section{Magnetically Confined Mountain Theory}
\label{sec:mountains_theory}
In this paper, we investigate the equilibrium of accreted matter on a white dwarf's (WD's) magnetic pole. The theory used here has its origins in neutron star physics \cite{payne2004burial,priymak2011quadrupole,rossetto2023magnetically}, but it can be successfully applied to white dwarfs. In this section, we summarize the main equations that govern the system and the techniques used to solve them. For a detailed discussion, see \citeauthor{rossetto2026magnetically}, \citeyear{rossetto2026magnetically} \cite{rossetto2026magnetically}.

We start by considering the WD's magnetic field $\mathbf{B}$ to be axisymmetric and poloidal. This allows us to write it as
\begin{equation}
    \label{mag_field}
    \mathbf{B} = \frac{\grad{\psi}}{r\sin\theta}\cross \hat{\phi},
\end{equation}
where $\psi=\psi(r,\theta)$ is the magnetic flux function, $(r,\theta,\phi)$ are spherical coordinates such that $\theta$ is the polar angle --- with respect to the magnetic axis --- and $\phi$ is the azimuthal angle. Then, we consider that the accreted matter, i.e. the mountain, obeys an isothermal equation of state $p = {c_{\rm s}}^2\rho$, with $p$ being the pressure, $c_{\rm s}$ the isothermal sound speed, and $\rho$ the density. Using these approximations in the MHD equations and imposing static equilibrium, we obtain
\begin{equation}
    \label{iso_grad_shafranov}
    \Delta^*\psi=-F'(\psi)e^{-(\Phi-\Phi_0)/c_s^2},
\end{equation}
where $\Phi = GM_*r/{R_*}^2$ is the surface approximation of the gravitational field, $G$ is Newton's constant, $M_*$ is the mass of the WD, $R_*$ its radius, $\Phi_0 = \Phi(R_*)$, $F(\psi)$ is a function that is determined by an ideal MHD condition (see equation \eqref{Fpsi}), and $\Delta^*$ is the operator
\begin{equation}
    \label{grad_shaf_op}
    \Delta^* = \frac{1}{4\pi r^2\sin^2\theta}\left[\pdv[2]{r}+\frac{\sin\theta}{r^2}\pdv{\theta}\left(\frac{1}{\sin\theta}\pdv{\theta}\right)\right].
\end{equation}
Equation \eqref{iso_grad_shafranov} is the Grad-Shafranov equation governing plasma equilibrium.

The mountain density profile $\rho = \rho(r,\theta)$ can be written in terms of the flux function $\psi$ as
\begin{equation}
    \label{rho}
    \rho = \frac{F(\psi)}{{c_s}^2}e^{-(\Phi-\Phi_0)/c_s^2}.
\end{equation}
In the case of ideal MHD, the magnetic field is frozen into the plasma. Using this condition together with equation \eqref{rho} we get
\begin{equation}
    \label{Fpsi}
    F(\psi) = \frac{{c_s}^2}{2\pi}\frac{dM}{d\psi}\left\{\int_\mathcal{C}ds\, r\sin\theta|\grad{\psi}|^{-1}e^{-(\Phi - \Phi_0)/{c_s}^2}\right\}^{-1},
\end{equation}
where $\dd{M}/\dd{\psi}$ is the mass-flux distribution. In our model, we choose a distribution that has the qualitative aspects of polar accretion, while maintaining it sufficiently smooth for numerical calculations,
\begin{equation}
    M(\psi) = \frac{M_{\rm a}}{2}\frac{1-e^{-b\psi/\psi_*}}{1 - e^{-b}},
    \label{eq:dmdpsi_pm04}
\end{equation}
with $M_{\rm a}$ being the accreted mass, $\psi_* = B_*{R_*}^2/2$, $b = R_{\rm m}/R_*$, and $R_{\rm m}$ being the radius of the magnetosphere. Equation \eqref{iso_grad_shafranov} has to be solved simultaneously with equation \eqref{Fpsi} for one to obtain a self-consistent solution. This is done by using an initial function $\psi_0$ and, then, calculating $F(\psi)$ using equation \eqref{Fpsi} and then solving equation \eqref{iso_grad_shafranov} for $\psi$. We repeat this process until $\psi$ converges to an equilibrium value. As for the boundary conditions for equation \eqref{iso_grad_shafranov}, we use $\psi(r,0) = 0$, $\psi(R_*,\theta) = \psi_{\rm d}(R_*,\theta)$ (where $\psi_{\rm d}$ is the dipolar field) and $\dd{\psi}/\dd{\theta} (r,\pi/2) = 0$. For the outer boundary $r=R_{\rm max}$, we impose that the value of the dipole moment does not change with $r$, which gives us
\begin{equation}
    \label{robin_bc}
    \pdv{\psi}{r} + \frac{\psi}{r} = 0.
\end{equation}
The Grad-Shafranov equation \eqref{iso_grad_shafranov} is solved using sparse solvers \cite{davis2004algorithm,virtanen2020scipy} on a discretised grid with log spacing of the $r$ coordinate, for better resolution near the surface.

\begin{figure*}
    \centering
    \begin{subfigure}[b]{0.49\textwidth}
        \centering
        \includegraphics[height=0.35\textheight]{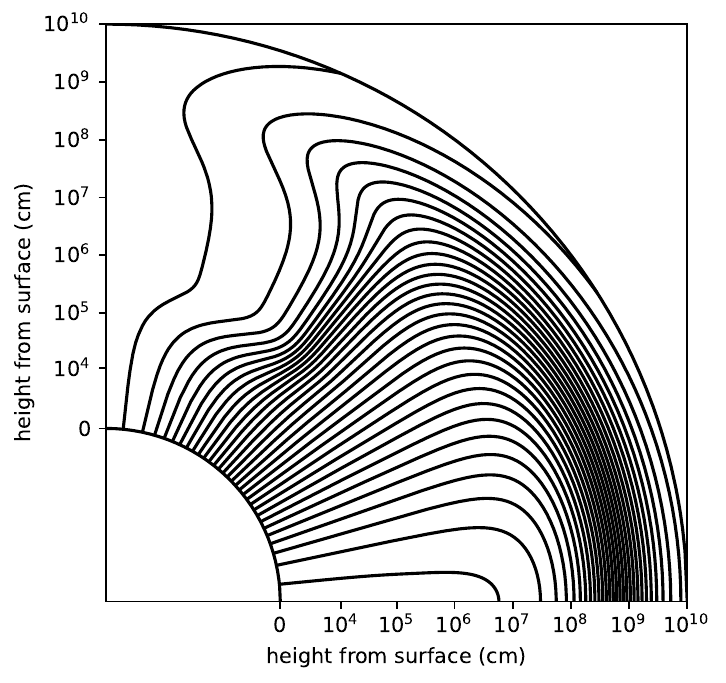}
        \caption{}
        \label{fig:psi_contours}
    \end{subfigure}
    \begin{subfigure}[b]{0.49\textwidth}
        \centering
        \includegraphics[height=0.35\textheight]{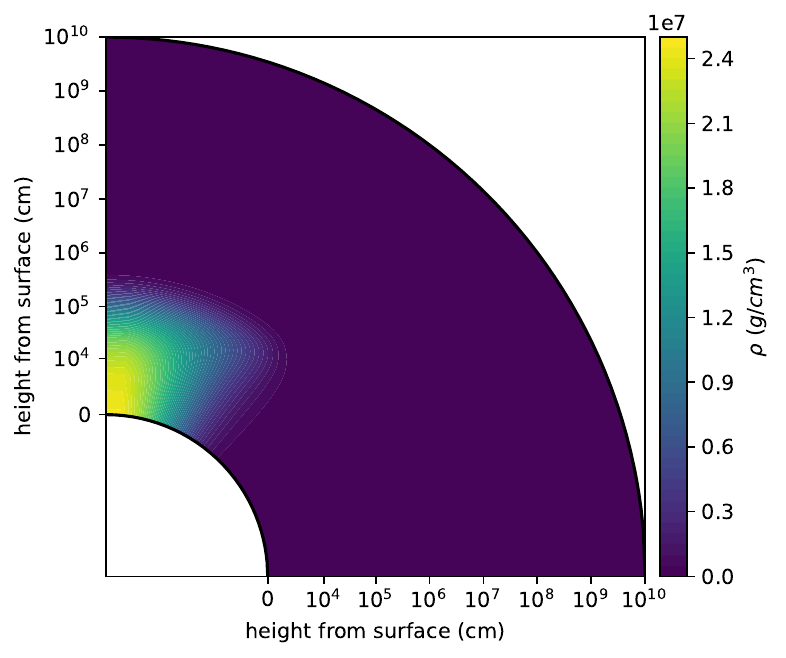}
        \caption{}
        \label{fig:density}
    \end{subfigure}
    \caption{Magnetic field deformation and mountain density profile for  $M_* = 0.8 M_\odot$, $R_* = 7\times10^8\,\mathrm{cm}$, $c_{\rm s} = 4.67\times10^{6}\,\mathrm{cm/s}$, $B_* = 5\times10^{7}\,\mathrm{G}$, $M_{\rm a} = 10^{-3}M_\odot$. (a) Magnetic field lines (isocontours of $\psi$). (b) Mountain density, $\rho$.}
    \label{fig:model_simulation}
\end{figure*}

Figure \ref{fig:model_simulation} shows an example output of our simulations, for the system's parameters $M_* = 0.8 M_\odot$, $R_* = 7\times10^8\,\mathrm{cm}$, $c_{\rm s} = 4.67\times10^{6}\,\mathrm{cm/s}$, $B_* = 5\times10^{7}\,\mathrm{G}$, $M_{\rm a} = 10^{-3}M_\odot$. Panel \ref{fig:psi_contours} shows the deformation of the magnetic field lines after the accreted mountain has spread equatorwards due to gravity, but was maintained in equilibrium by the magnetic tension. Panel \ref{fig:density} shows the density profile of the magnetically confined mountain. This simulation has converged, i.e., $\psi$ has reached an equilibrium value, and the total mass has varied by no more than a few percent throughout the self-consistent solution scheme.

\section{Gravitational Wave Emission and Ellipticity}
\label{sec:gw_model}
We explore a gravitational wave (GW) emission mechanism in rapidly rotating magnetic white dwarfs (WDs), arising from magnetically confined mountains. In our scenario, the magnetic poles and the rotation axis are non-aligned, and the WD accretes matter from a companion star through magnetically funneled accretion — either via a magnetically truncated disk or a diskless accretion stream \cite{welsh1998doppler,warner1995cataclysmic}. Because accreted matter is channeled exclusively to the magnetic poles, the WD develops a non-axisymmetric shape, which is the key ingredient for GW emission. In Section \ref{sub-sec:gw_model}, we summarize the GW emission model for a rigidly rotating body with non-zero ellipticity. Then, in Section \ref{sub-sec:max_ellip}, we use the theory of magnetically confined mountains, presented in Section \ref{sec:mountains_theory}, to calculate the maximum ellipticity that accreting WDs can attain as a function of their magnetic field strength and temperature.

\subsection{Gravitational Wave Emission Model}
\label{sub-sec:gw_model}

We model the WD in a co-rotating body frame $(x_{1}, x_{2}, x_{3})$ aligned with its principal axes of inertia $(I_{1}, I_{2}, I_{3})$. The star rotates with an angular velocity $\Omega$ ($2 \pi f$) about the $x'_{3}$ axis of an inertial reference frame $(x'_{1}, x'_{2}, x'_{3})$. We align the magnetic dipole with the $x_3$ axis, such that the magnetic and rotation axes are misaligned by an angle $\alpha$. Assuming axial symmetry ($I_{1} = I_{2}$), the strain amplitude of the emitted GWs measured at a distance $d$ is given by \cite{jaranowski1998data,shapiro1983black,maggiore2008gravitational}
\begin{equation}
h_{0} = \frac{16 \pi^2 G}{c^{4}} \frac{I_0 \varepsilon f^{2}}{d}  \sin ^{2}\alpha,
\label{AmpAC2}
\end{equation}
where $I_0 = (2/5)M_*{R_*}^2$ is the pre-accretion moment of inertia of the star, and $\varepsilon$ is the star's ellipticity, defined as
\begin{equation}
    \label{epsilon_I}
    \varepsilon = \frac{I_2-I_3}{I_0}.
\end{equation}
Using the standard definition of the moment of inertia as the integral of the mass density times the square distance to the rotating axis, we can write the ellipticity as 
\begin{equation}
    \label{epsilon}
    \varepsilon=\frac{1}{2I_0}\int_{V'} \rho(r,\theta) r^2(3\cos^2\theta-1)r^2\sin\theta\dd{r}\dd{\theta}\dd{\phi},
\end{equation}
where $\rho$ is the mass density, $V'$ is the volume of the body and $(r,\theta, \phi)$ are the standard spherical coordinates in the body frame -- with $\theta$ being the polar angle with respect to the magnetic axis and $\phi$ being the azimuthal angle around it \cite{vigelius2009improved}. 

\begin{figure}
    \centering
    \includegraphics[width=0.49
    \textwidth]{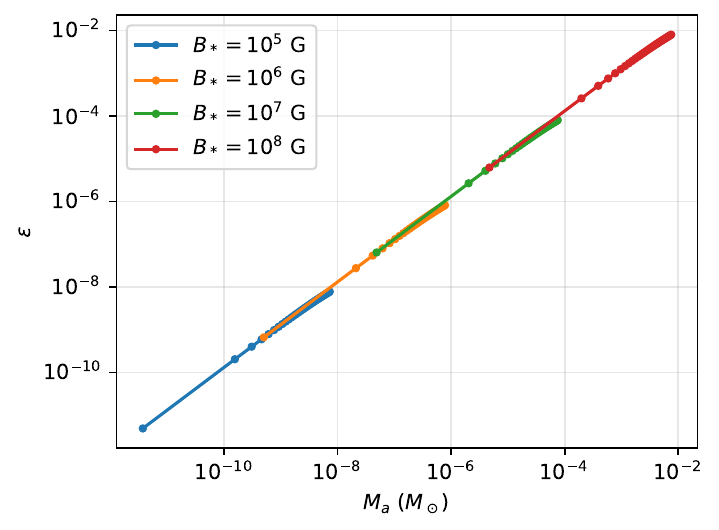}
    \caption{White dwarf ellipticity $\varepsilon$ per accreted mass $M_{\rm a}$ for four different values of magnetic field $B_*$. For all the simulations shown, $M_* = 0.8 M_\odot$, $R_* = 7\times10^8\,\mathrm{cm}$, $T = 1.58\times10^{5}\,\mathrm{K}$ and $b=10$.}
    \label{fig:ellip_per_Ma}
\end{figure}

Equation (\ref{AmpAC2}) assumes an optimal viewing geometry where the observer's line of sight is parallel to the rotation axis ($x'_{3}$), an orientation that maximizes the detectable GW amplitude \cite{maggiore2008gravitational}. We can also parameterize Equation (\ref{AmpAC2}) using characteristic astrophysical values. Scaling the amplitude to typical WD parameters, we obtain for $\alpha = 45^\circ$
\begin{align}
    &h_0 \approx 6.0 \times10^{-26}
    \left(\frac{M_*}{0.8M_\odot}\right)
    \left(\frac{R_*}{7\times10^8\,\mathrm{cm}}\right)^2 \label{AmpACC}
    \\
    &\left(\frac{f}{0.03\,\mathrm{Hz}}\right)^2
    \left(\frac{\varepsilon}{10^{-5}}\right)
    \left(\frac{100\,\mathrm{pc}}{d}\right). \nonumber
\end{align}
This relation quantifies the GW emission from a WD that hosts a non-axisymmetric deformation, i.e., a mountain. The resulting amplitude is governed by the height and mass of the magnetically confined mountain, expressed through the ellipticity, while growing with the square of the rotation frequency and decaying linearly with the observer's distance. In Figure \ref{fig:ellip_per_Ma}, we show the dependency of the ellipticity $\varepsilon$ per accreted mass $M_{\rm a}$ for several Grad-Shafranov equilibria and for selected values of the star's magnetic field $B_*$. Consistent with equation \eqref{epsilon}, $\varepsilon$ increases almost linearly with $M_{\rm a}$. Nevertheless, the maximum amount of mass that can be supported in equilibrium is dependent on the value of the magnetic field.

\subsection{Maximum Ellipticity}
\label{sub-sec:max_ellip}

The extent of the magnetic field deformation relies fundamentally on the accreted mass, $M_{\rm a}$, channeled to the magnetic poles. In the low accreted mass regime, magnetic tension vastly dominates over local pressure and gravitational forces, allowing the accreted mountain to grow without perturbing the underlying dipolar magnetic topology. Conversely, as $M_{\rm a}$ increases, the gravitational pull forces the mountain to spread equatorward, dragging the frozen-in magnetic field lines along with the plasma. This distorted equilibrium state is dictated by the balance between pressure, magnetic and gravitational forces, as formally described by the Grad-Shafranov equation (see Section~\ref{sec:mountains_theory}).

However, as the accreted mass approaches a critical threshold, the magnetic field lines experience severe distortion that ultimately triggers magnetic reconnection and the formation of magnetic islands. This topological breakdown signifies a macroscopic loss of equilibrium, where our numerical simulations no longer converge. We investigate this critical threshold across a parameter space representative of magnetic WDs. Throughout our simulations, we adopt fiducial values for the stellar mass and radius of $M_* = 0.8\, M_\odot$ and $R_* = 7\times10^8\,\mathrm{cm}$, respectively. For a discussion on the effects of the mass-radius values on the ellipticity and on the general relativistic corrections; see Appendix \ref{app:compactness}. Furthermore, we consider polar magnetic field strengths in the range $1.00\times10^5 \, \mathrm{G} \leq B \leq 3.00 \times 10^8 \, \mathrm{G}$ \cite{dufour2017montreal, amorim2023catalog,schwope2025polarcat}, and the accreted mountain is modeled as an isothermal perfect fluid, with temperatures ranging from $2.00 \times 10^{4} \,\mathrm{K} \leq T \leq 1.00 \times 10^{7} \,\mathrm{K}$. The lower bound corresponds to typical white dwarf surface temperatures \cite{dufour2017montreal, amorim2023catalog, pala2017effective}, whereas the upper limit accounts for shock-heating within the accretion column \cite{saxton2005twotemperature,jose2025hydrodynamic,yuasa2010white}. In our formulation, the temperature determines the isothermal sound speed, given by $c_s=\sqrt{(k_B T)/(\mu m_H)}$, where $k_B$ is the Boltzmann constant, $\mu = 2$ is the mean molecular weight, and $m_H$ is the hydrogen mass.

\begin{figure*}
    \centering
    \begin{subfigure}[b]{0.49\textwidth}
        \centering
        \includegraphics[width=\textwidth]{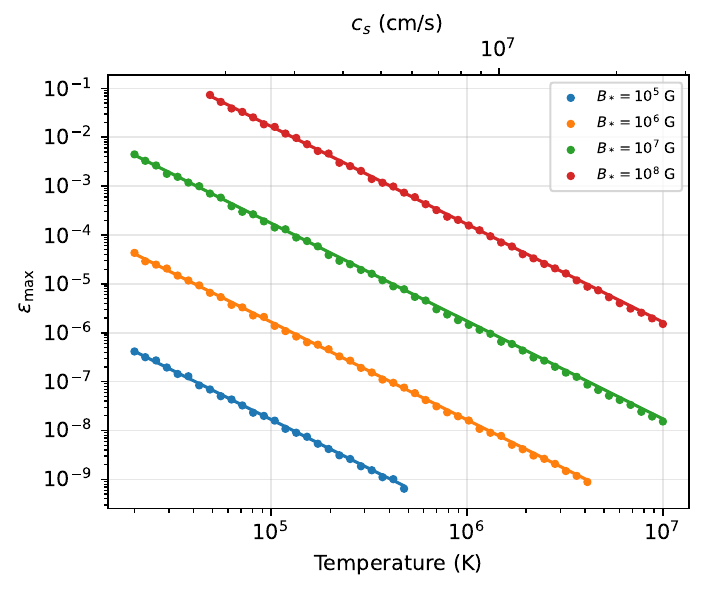}
        \caption{}
        \label{fig:max_ellip_T}
    \end{subfigure}
    \begin{subfigure}[b]{0.49\textwidth}
        \centering
        \includegraphics[width=\textwidth]{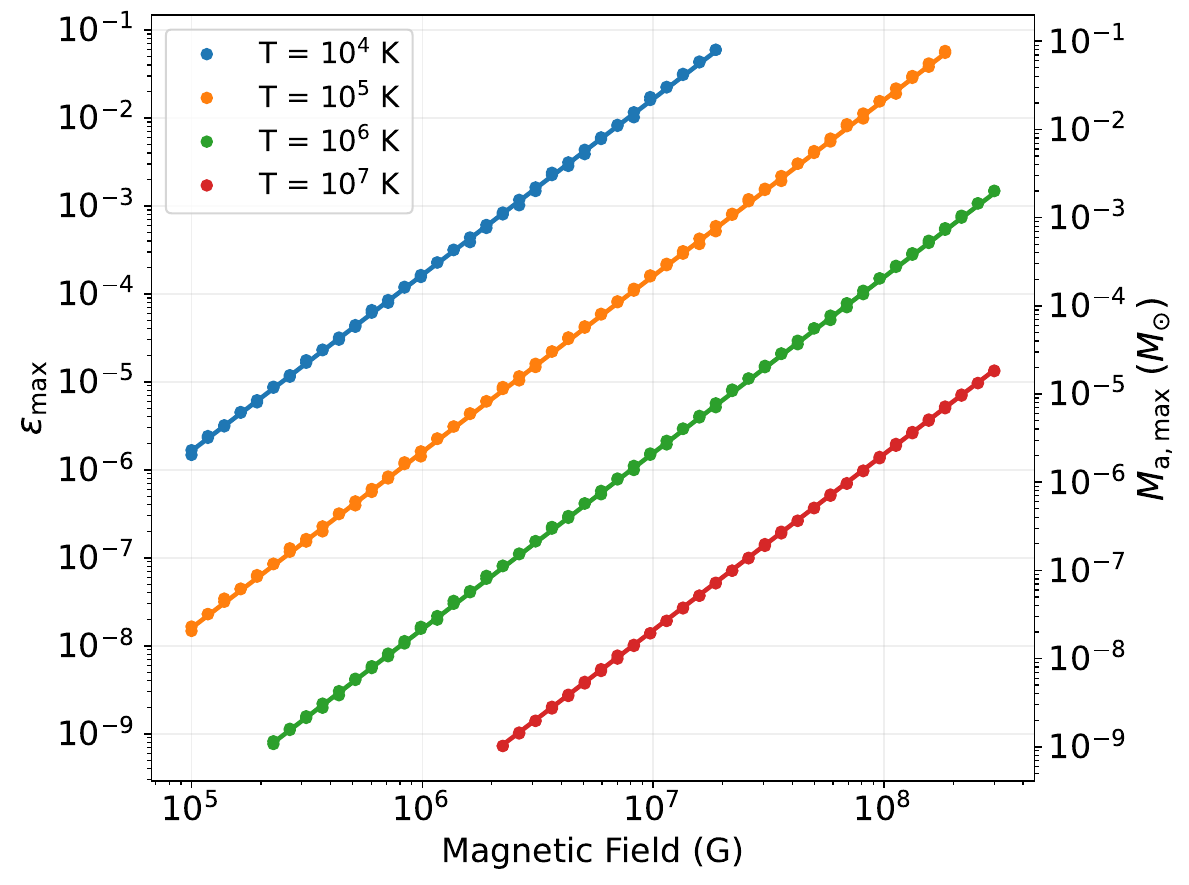}
        \caption{}
        \label{fig:max_ellip_B}
    \end{subfigure}
    \caption{Ellipticity $\varepsilon_{\rm max}$ at maximum accreted mass $M_{\rm a, max}$ as a function of magnetic field and temperature. For all the simulations shown here: $M_* = 0.8\,M_\odot$, $R_* = 7 \times 10^8\,\mathrm{cm}$ and $b=10$. (a) Temperature dependence, for four fixed values of magnetic field. (b) Magnetic field dependence, for four fixed values of temperature.}
    \label{fig:max_ellip}
\end{figure*}

To precisely determine the $M_{\rm a, max}$ permitted by the Grad-Shafranov equilibrium under these conditions, we employ a binary search algorithm. This routine searches for the critical mass between a lower bound of $10^{-9} M_\odot$ (where the numerical solution consistently converges) and an upper bound of $10^{-1} M_\odot$ (where convergence fails). For a comprehensive description of the algorithm's implementation and operation, we refer the reader to \citeauthor{rossetto2026magnetically} (\citeyear{rossetto2026magnetically}) \cite{rossetto2026magnetically}. At the maximum accreted mass value, we take the corresponding ellipticity value, $\varepsilon_{\rm max}$, and verify how it behaves with the system's parameters. Figure \ref{fig:max_ellip} shows the results of our investigations. Panel \ref{fig:max_ellip_T} shows the dependency on the temperature $T$ --- equivalently, the sound speed $c_{\rm s}$ --- while Panel \ref{fig:max_ellip_B} shows the dependency on the magnetic field $B_*$. The lines joining the points are power-law fits, that can be combined in the following expression
\begin{equation}
    \label{max_mass_formula_fit}
    \varepsilon_{\rm max} = (4.0\pm0.2)\times10^{-3}
    \left(\frac{T}{10^5 \, \mathrm{K}}\right)^{-2}\left(\frac{B_*}{5\times10^7 \, \mathrm{G}}\right)^{2},
\end{equation}
which is the same functional dependency found in Rossetto et al. (2026) \cite{rossetto2026magnetically} for $M_{\rm a, max}$, but with a different multiplicative coefficient.

As observed, for a given temperature, both $M_{\rm a, max}$ and $\varepsilon_{\rm max}$ scale with the magnetic field following a power-law relationship. This behavior is physically expected: a stronger magnetic field provides greater magnetic tension, which more effectively resists the deformation of the field lines. Consequently, the magnetic field can sustain a more massive accreted mountain, yielding a higher ellipticity. Conversely, for a fixed magnetic field strength, the maximum sustainable mass and ellipticity decrease as the temperature rises. This inverse correlation arises because higher temperatures translate to increased thermal energy and a higher local sound speed $c_s$. Due to the ideal MHD flux-freezing condition, this enhanced thermal pressure gradient more readily overcomes the magnetic tension, severely dragging and distorting the field lines. This early onset of magnetic deformation limits the amount of matter that can be stably confined, thereby preventing the accumulation of a larger mountain.

In summary, our parameter space exploration delineates the physical boundaries of magnetic confinement in these systems. The derived values of $\varepsilon_{\rm max}$ represent the most extreme structural deformation the white dwarf can sustain before the onset of macroscopic instabilities. Establishing these physical upper limits provides the essential foundation required to estimate the maximum GW emission generated by the accreted mountain and probe the prospects of its detectability.

\section{Estimates and Detectability of Continuous Gravitational Waves}
\label{sec:results}
Building upon the maximum sustainable ellipticities, $\varepsilon_{\rm max}$, established in the previous section, we now evaluate the resulting CW emission. Adopting the fiducial WD parameters, we place the system at a reference distance of $d = 100$~pc and assume a rotation period of $P = 30$~s. Utilizing Equation~\ref{AmpAC2}, we compute the upper limits of the GW amplitude, $h_{\rm 0,max}$, generated by the accreted mountain at the magnetic poles.

\begin{figure}
    \centering
    \includegraphics[width=0.49
    \textwidth]{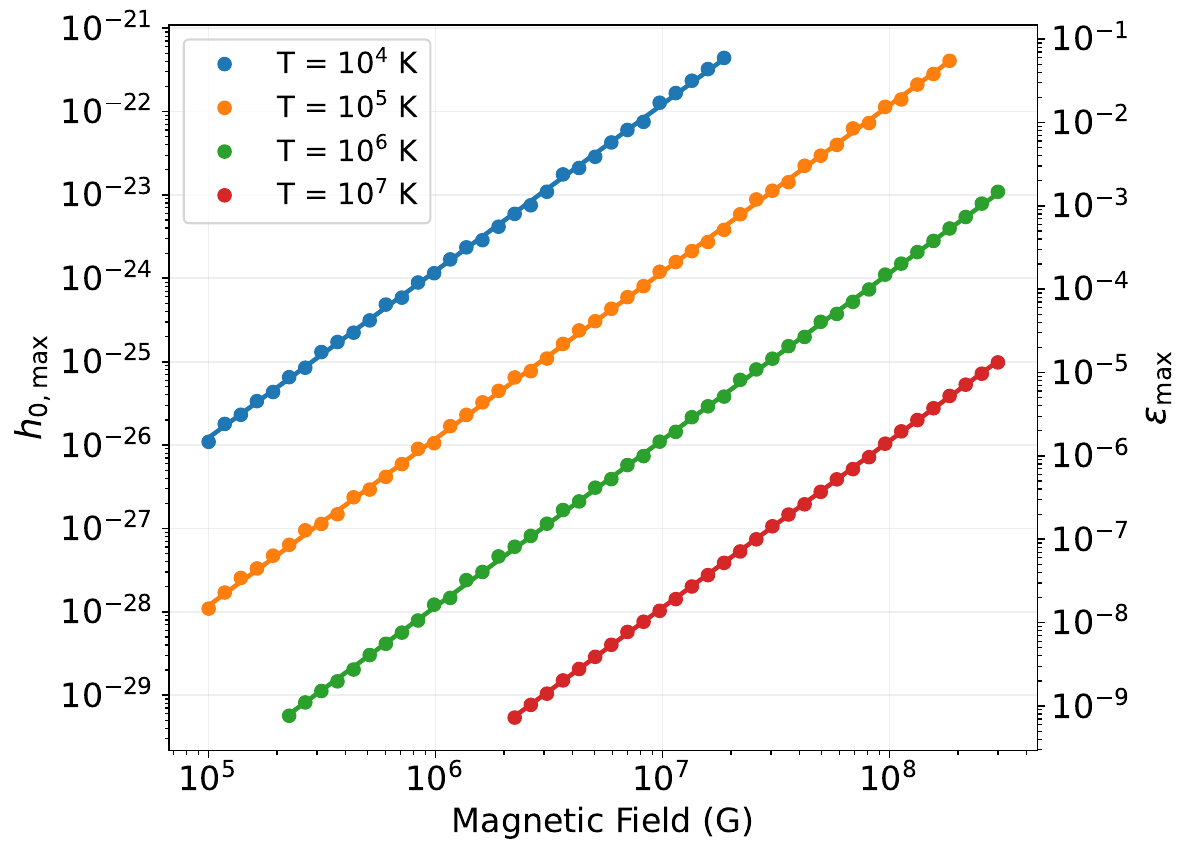}
    \caption{Maximum GW amplitude $h_{\rm 0, max}$, and corresponding maximum ellipticity $\varepsilon_{\rm max}$, as functions of the polar magnetic field strength, $B_*$. The curves denote four distinct isothermal temperatures for the accreted mountain. All models adopt fiducial parameters of $M_* = 0.8\, M_\odot$, $R_* = 7 \times 10^8\,\mathrm{cm}$, $P = 30$~s and $d = 100$~pc.}
    \label{fig:ellip_and_h_per_B}
\end{figure}

The results of this estimation are presented in Figure~\ref{fig:ellip_and_h_per_B}, which illustrates $h_{\rm 0,max}$ as a function of the polar magnetic field, $B_*$, across four isothermal mountain temperatures $T$. Because the GW amplitude is directly proportional to the ellipticity, $h_{\rm 0,max}$ follows the behavior observed in Figure~\ref{fig:max_ellip_B}: the amplitude scales with the magnetic field strength and decreases at higher temperatures. Consequently, the numerical data can be well parameterized by a power-law fit, yielding the following relation:
\begin{equation}
    \label{h_max_fit}
    h_{\rm 0, max} = (3.0\pm0.2)\times10^{-23}\left(\frac{T}{10^5 \, \mathrm{K}}\right)^{-2}\left(\frac{B_*}{5\times10^7 \, \mathrm{G}}\right)^{2}.
\end{equation}
As physically expected, this scaling relation exhibits the same functional dependence found for the maximum ellipticity (Equation~\ref{max_mass_formula_fit}). In fact, this expression can be analytically recovered by substituting Equation~\ref{max_mass_formula_fit} into Equation~\ref{AmpACC}.

The predicted GW amplitudes span a broad range, from negligible values ($\sim 10^{-29}$) up to $\sim 10^{-21}$, placing the configurations with higher ellipticity within the sensitivity bands of planned space-based interferometers. Notably, Figure~\ref{fig:ellip_and_h_per_B} highlights an important degeneracy of parameters: while the definition of a maximum ellipticity inherently determines the required accreted mass, this state can be achieved through multiple combinations of thermal and magnetic field properties. Nevertheless, reaching the highest ellipticity regimes demands specific combinations of $B_*$ and $T$ capable of providing sufficient magnetic confinement, as previously demonstrated in Figure~\ref{fig:max_ellip_B}.

\begin{figure}
    \centering
    \includegraphics[width=0.49
    \textwidth]{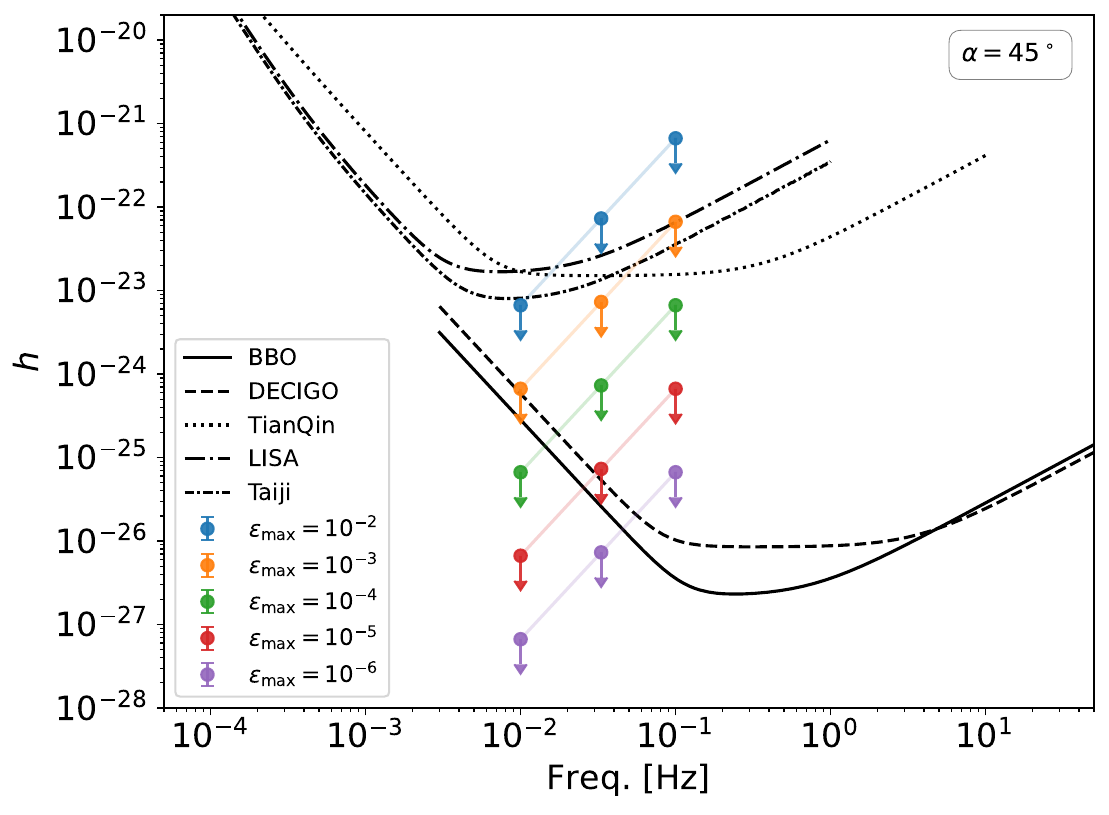}
    \caption{Upper limits of the GW amplitude versus rotation frequency for varying maximum ellipticities ($\varepsilon_{\rm max}$) rotation periods in the range $10\, \mathrm{s} \leq P \leq 100\, \mathrm{s}$. The markers specifically denote the periods $P = \{10, 30, 100\}$~s. The sensitivity curves of the space-based detectors LISA, Taiji, TianQin, BBO, and DECIGO are shown assuming an $\mathrm{SNR} = 8$ and an observation time of $\tau_{\rm obs} = 1$~yr. These curves are shifted in frequency to properly map the $f_{\rm GW} = 2f$ emission relation onto the rotational frequency axis. The models adopt an inclination angle of $\alpha = 45^{\circ}$, a source distance of $d = 100\,\mathrm{pc}$, and the fiducial WD parameters $M_* = 0.8\, M_\odot$ and $R_* = 7 \times 10^8$~cm.}
    \label{fig:h_freq}
\end{figure}

To assess the observational prospects, we then selected five reference ellipticities, 
$\varepsilon_{\rm max} = (10^{-2}, \, 10^{-3}, \, 10^{-4}, \, 10^{-5}, \, 10^{-6})$, and rotation periods in the range $10\, \mathrm{s} \leq P \leq 100\, \mathrm{s}$, which are consistent with the population of fast-spinning WDs \cite{terada2008suzaku,mereghetti2021new,pelisoli2021found,pelisoli202353minperiod}. Assuming an inclination angle of $\alpha = 45^{\circ}$, Figure~\ref{fig:h_freq} displays the upper limits of the GW amplitude as a function of the GW frequency. We compare these signals against the theoretical sensitivity curves of future space-based observatories, namely LISA, TianQin, Taiji, BBO, and DECIGO \cite{amaro-seoane2017laser,harry2006laser,yagi2011detector,ruan2020taiji}. These instrumental noise curves are expressed in terms of the minimum detectable amplitude, $h_{\rm min}$, which can be measured by the detector for a periodic signal, considering a given signal-to-noise ratio (SNR) and an observation time $\tau_{\rm obs}$ \cite{maggiore2008gravitational}. Accordingly, the sensitivity curves are evaluated at SNR = $8$ and $\tau_{\rm obs} = 1$~yr. Because the $x$-axis represents the stellar rotation frequency rather than the GW frequency ($f_{\rm GW} = 2f$), the instrumental sensitivity curves have been appropriately shifted by a factor of two along the frequency axis.

Analyzing the detectability horizons in Figure~\ref{fig:h_freq}, we find that LISA, Taiji, and TianQin are primarily sensitive to systems with higher ellipticities. Specifically, these instruments could detect WDs harboring an ellipticity of $\sim 10^{-2}$ provided their rotation periods are shorter than $\sim 50$~s. TianQin exhibits slightly broader prospects, potentially detecting ellipticities of $\sim 10^{-3}$ if $P \lesssim 20$~s. Physically, these configurations demand a highly magnetized and relatively cool accreted mountain ($T \lesssim 10^{5}$~K and $B_* \gtrsim 10^{7}$~G) to maintain stability against the breaking of the field lines.

The proposed next-generation interferometers BBO and DECIGO offer significantly higher sensitivities, capable of observing lower ellipticities down to $\varepsilon \sim 10^{-6}$, provided the star spins sufficiently fast ($P \lesssim 20$~s). This expanded sensitivity opens up a wider parameter space of detectable ($T, B_*$) configurations. Nonetheless, the physical constraints remain: if the mountain is relatively cold ($T \lesssim  10^{4}$~K), magnetic fields as low as $B_* \sim 10^{5}$~G can support a detectable $\sim 10^{-6}$ ellipticity. However, as the temperature approaches shock-heated values ($T \sim 10^{7}$~K), the intense thermal pressure mandates much stronger magnetic fields ($B_* \gtrsim 10^{8}$~G) to prevent loss of confinement of the accreted matter.

\begin{figure*}
    \centering
    \includegraphics[width=0.99
    \textwidth]{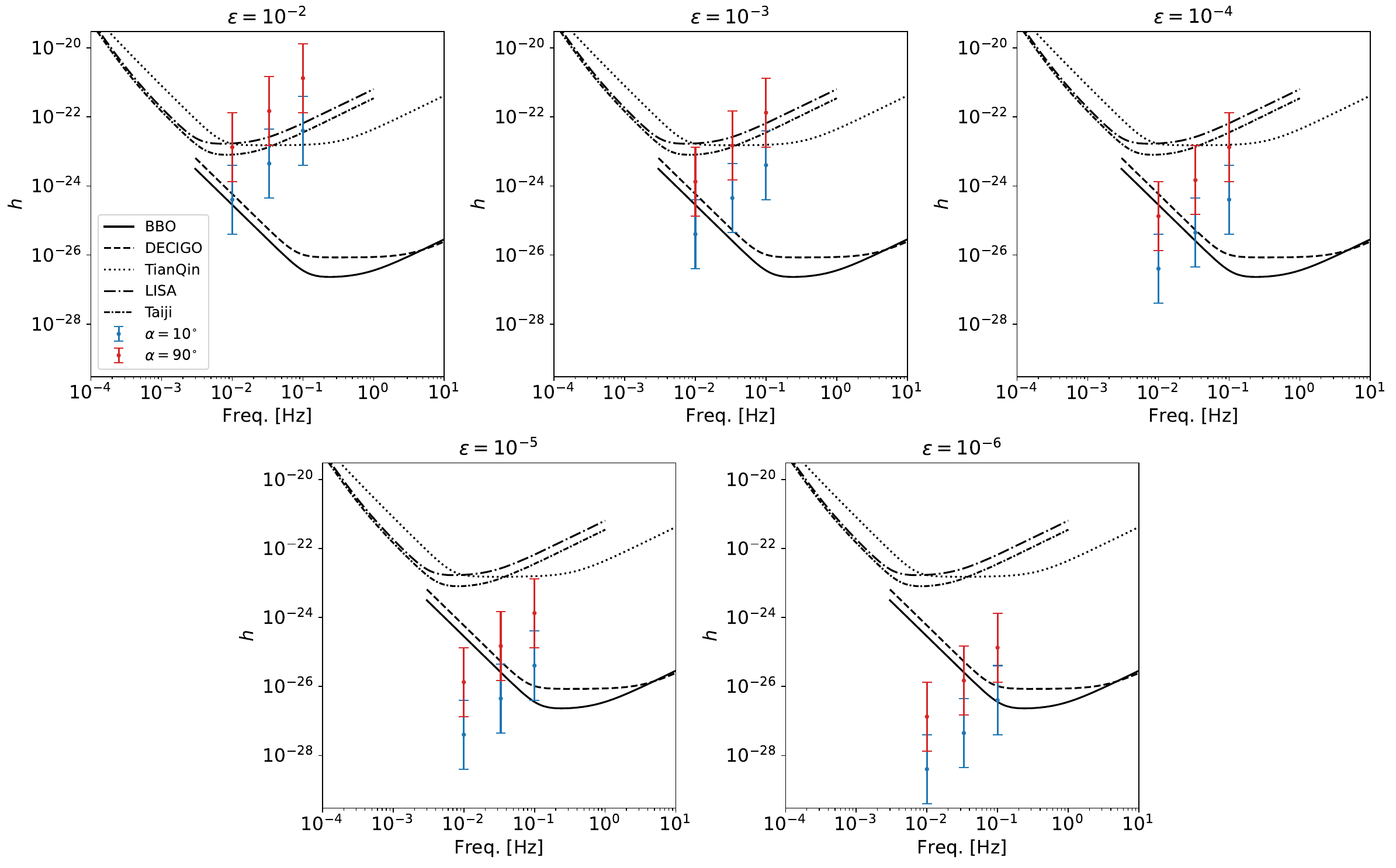}
    \caption{GW amplitude for different distance values taking into account $\alpha = 90^{\circ}$  and $\alpha = 10^{\circ}$. The vertical bars stands for  $10 {\rm pc} \leq d \leq 1 {\rm kpc}$, from top to bottom. The different panels represent the ellipticity $\varepsilon$ created by the accreted mass at the WD's magnetic poles. The sensitivity curves of the space-based detectors LISA, TianQin, Taiji, BBO and DECIGO are shown considering an SNR $= 8$ and $\tau_{\rm obs} = 1$~yr.}
    \label{fig:h_freq_dist}
\end{figure*}

To further evaluate the observational prospects of these sources, we investigate the dependence of the GW amplitude on the inclination angle, $\alpha$, and the source distance, $d$. We compute the amplitudes for a distance range of $10 \, {\rm pc} \leq d \leq 1 \, {\rm kpc}$, consistent with observations of fast-spinning WDs \cite{terada2008suzaku, pelisoli2021found,gaiacollaboration2023gaia}. Additionally, we consider two representative inclination angles: $\alpha = 10^\circ$, corresponding to a nearly aligned rotator, and $\alpha = 90^\circ$, an orthogonal rotator that maximizes the GW emission.

Figure \ref{fig:h_freq_dist} presents the results of this analysis across five panels, each corresponding to a distinct ellipticity, $\varepsilon$. In each panel, vertical bars indicate the GW amplitude variation over the specified distance range (with the upper and lower bounds of the bars corresponding to $d = 10$~pc and $1$~kpc, respectively) for the three selected rotation periods ($P = (10, \, 30, \, 100)$~s). The results for both inclination angles are shown for each period. As in the previous analysis, these calculated amplitudes are evaluated against the instrumental sensitivities of LISA, Taiji, TianQin, BBO, and DECIGO, assuming $\mathrm{SNR} = 8$ and $\tau_{\rm obs} = 1$~yr.

Focusing first on the LISA, Taiji, and TianQin observatories, detection is feasible for WDs with ellipticities $\varepsilon \gtrsim 10^{-3}$, subject to the specific system parameters. At $\varepsilon = 10^{-2}$, configurations with $P \lesssim 30$~s and $\alpha \sim 90^\circ$ fall within the observable range of LISA and TianQin across the entire $1$~kpc distance span. Under the same ellipticity, Taiji can detect sources with smaller inclination angles, provided the rotation periods are shorter. For an ellipticity of $\varepsilon = 10^{-3}$, these sources remain detectable by Taiji up to distances of $d \sim 1$~kpc, provided the system is a rapidly spinning ($P \sim 10$~s), nearly orthogonal rotator ($\alpha \sim 90^\circ$). For a lower ellipticity of $10^{-4}$, the signals fall below the detection thresholds of LISA and TianQin. In this regime, detection is restricted to Taiji and requires a specific parameter combination, namely $\alpha \sim 90^\circ$, $P \sim 10$~s, and a distance of $d \lesssim 100$~pc.

Turning to the proposed next-generation observatories, BBO and DECIGO, WDs with ellipticities $\varepsilon \gtrsim 10^{-3}$ and rotation periods $P \lesssim 30$~s are detectable across the entire $1$~kpc distance range. For slower rotators ($P \sim 100$~s), successful observation necessitates large inclination angles ($\alpha \sim 90^\circ$) or close proximity to the observer ($d \lesssim 100$~pc). At intermediate ellipticities, $\varepsilon = (10^{-4}, \, 10^{-5})$, the full distance range remains accessible only if the star is spinning very rapidly ($P \sim 10$~s). Systems with $P \sim 30$~s in this ellipticity regime are observable solely under restricted combinations of inclination and distance, while those with $P \sim 100$~s drop below the instrumental capabilities. For the lowest considered ellipticity ($\varepsilon = 10^{-6}$), the signals are measurable exclusively for very fast-spinning sources presenting large inclination angles or located at shorter distances. Finally, we reiterate that reaching the higher ellipticity values required for wider detectability relies on restricted combinations of temperature and magnetic field strength, which are necessary to preserve the magnetohydrodynamic equilibrium of the accreted mountain.

\begin{table} \centering 
\begin{threeparttable}
    \caption{Fast-spinning magnetic WDs along with their main parameters: Period (P), magnetic field ($B_*$) and distance to source (d).} 
    \label{tab:Tab_Param} 
    
    \begin{tabular}{lccc} \hline \hline
        \textbf{SYSTEMS} & \textbf{\begin{tabular}[c]{@{}c@{}}P\\ (s)\end{tabular}} & \textbf{\begin{tabular}[c]{@{}c@{}}$B_*$ \\ (G)\end{tabular}} & \textbf{\begin{tabular}[c]{@{}c@{}}d\\ (pc)\end{tabular}} \\ \hline 
        LAMOST J0240+1952\tnote{a} & 24.93 & $10^{7}$           & 592.98\tnote{e} \\ 
        CTCV J2056-3014\tnote{b}   & 29.61 & $10^{6}$           & 256.29\tnote{e} \\ 
        AE Aquarii\tnote{c}        & 33.08 & $5.0 \times 10^{7}$ & 91.87\tnote{e}  \\ 
        V1460 Herculis\tnote{d}    & 38.87 & $10^{7}$           & 266.03\tnote{e} \\ \hline \hline
    \end{tabular} 
    $^a$ \cite{pelisoli2021found};  $^b$ \cite{lopesdeoliveira2020ctcv,salcedo2024broadband}; $^c$ \cite{terada2008suzaku,kitaguchi2014nustar}; $^d$ \cite{ashley2020v1460,kjurkchieva2017newly}; $^e$ \cite{gaiacollaboration2023gaia}.

\end{threeparttable}
\end{table}

To evaluate the observational prospects of actual astrophysical systems, we apply our formalism to four known fast-spinning magnetic WDs. The fundamental properties of these sources are summarized in Table~\ref{tab:Tab_Param}. All selected sources reside in binary systems and are classified as magnetic cataclysmic variables hosting rapidly rotating WDs, implying they have gained angular momentum via mass accretion from their companions \cite{beskrovnaya2015origin,pelisoli2021found}. As the magnetic field strengths for LAMOST~J0240+1952 and V1460~Her remain observationally unconstrained, we adopt a fiducial value of $B_* = 10^{7}$~G. This represents a typical field strength for magnetic cataclysmic variables (see, e.g., Figure 30 in \cite{ferrario2020magnetic}) and is insufficient for the magnetic interaction with the secondary star to lock the WD's rotation in synchronicity with the orbital period. For CTCV~J2056-3014, we adopt a magnetic field estimate of $B_* = 10^{6}$~G derived by \citeauthor{salcedo2024broadband} (\citeyear{salcedo2024broadband}) \cite{salcedo2024broadband}, although we note that alternative studies suggest a lower field strength for this WD \cite{lopesdeoliveira2020ctcv}.

\begin{figure}
    \centering
    \includegraphics[width=0.49
    \textwidth]{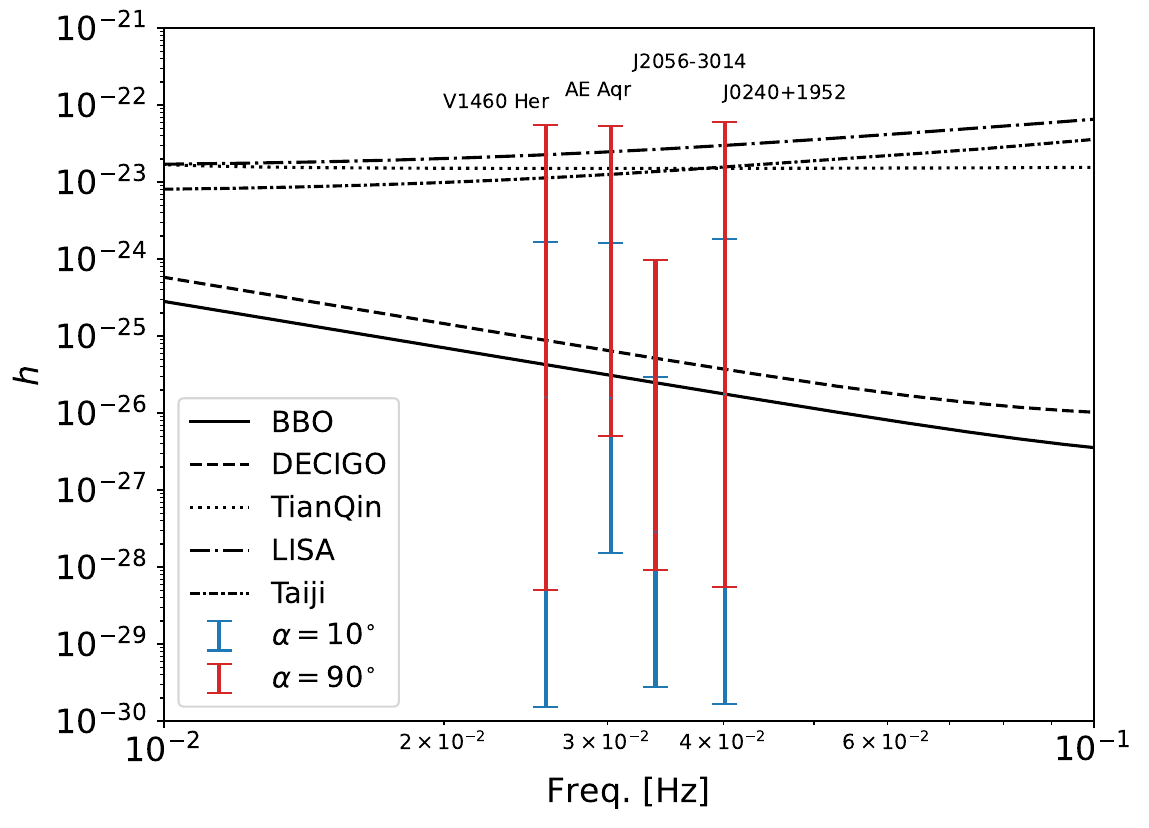}
    \caption{GW amplitude for fast-spinning magnetic white dwarfs, evaluated at inclination angles of $\alpha = 90^\circ$ and $\alpha = 10^\circ$. The vertical bars span the amplitude variation associated with the temperature range $T_{\rm min} \leq T_{\rm WD} \leq T_{\rm max}$ (from top to bottom, respectively). These temperature limits depend on the specific magnetic field strength of the WD. The sensitivity bands for the LISA, Taiji, TianQin, BBO, and DECIGO space-based detectors are shown assuming SNR $= 8$ and $\tau_{\rm obs} = 1$~yr.}
    \label{fig:h_freq_sources}
\end{figure}

Using the parameters from Table~\ref{tab:Tab_Param}, we calculate the maximum GW amplitudes across the previously defined temperature range for the accreted mountain. The results of this analysis are detailed in Table~\ref{tab:Tab_source_ampl} and illustrated in Figure~\ref{fig:h_freq_sources}. For the LAMOST~J0240+1952 and V1460~Her systems, the $\varepsilon_{\rm max}$ and $h_{\rm 0,max}$ are computed considering the entire temperature domain ($10^4 \leq T \leq 10^7$~K), since the accreted mountain maintains magnetohydrodynamic equilibrium under the assumed magnetic field across these temperatures. Conversely, for AE~Aquarii and CTCV~J2056-3014, the evaluated temperature ranges are restricted. We exclude temperatures below $10^5$~K and above $10^7$~K, respectively, because such specific $(T, B_*)$ lie outside our search limiting parameters.

\begin{table*}
\centering
    \caption{Derived physical parameters and expected gravitational wave amplitudes for the selected known fast-spinning white dwarfs. The table presents the assumed polar magnetic field ($B_*$) and temperature ($T$) range for the accreted mountain, along with the calculated maximum accreted mass ($M_{\rm a, max}$) and maximum ellipticity ($\varepsilon_{\rm max}$). The resulting upper limits for the GW amplitude ($h_{\rm 0, max}$) are evaluated at inclination angles of $\alpha = 10^\circ$ and $\alpha = 90^\circ$.} 
    \label{tab:Tab_source_ampl}
    
    \begin{tabular}{cccccc}
        \hline\hline
        \textbf{\begin{tabular}[c]{@{}c@{}}Temperatures\\ (K)\end{tabular}} & \textbf{\begin{tabular}[c]{@{}c@{}}$B_*$\\ (G)\end{tabular}} & \textbf{\begin{tabular}[c]{@{}c@{}}$M_{\rm a, max}$ \\ ($M_\odot$)\end{tabular}} & \textbf{$\varepsilon_{\rm max}$} & \textbf{\begin{tabular}[c]{@{}c@{}}$h_{\rm 0, max}$\\ ($\alpha = 90^\circ$)\end{tabular}} & \textbf{\begin{tabular}[c]{@{}c@{}}$h_{\rm 0, max}$\\ ($\alpha = 10^\circ$)\end{tabular}} \\ \hline
        \multicolumn{6}{c}{LAMOST J0240+1952} \\ \hline
        $10^4$ & $10^7$ & $2.3 \times 10^{-2}$ & $1.7 \times 10^{-2}$ & $6.0 \times 10^{-23}$ & $1.8 \times 10^{-24}$ \\
        $10^5$ & $10^7$ & $2.3 \times 10^{-4}$ & $1.6 \times 10^{-4}$ & $5.8 \times 10^{-25}$ & $1.8 \times 10^{-26}$ \\
        $10^6$ & $10^7$ & $2.2 \times 10^{-6}$ & $1.6 \times 10^{-6}$ & $5.6 \times 10^{-27}$ & $1.7 \times 10^{-28}$ \\
        $10^7$ & $10^7$ & $2.1 \times 10^{-8}$ & $1.5 \times 10^{-8}$ & $5.5 \times 10^{-29}$ & $1.6 \times 10^{-30}$ \\ \hline
        \multicolumn{6}{c}{CTCV J2056-3014} \\ \hline
        $10^4$ & $10^6$ & $2.4 \times 10^{-4}$ & $1.7 \times 10^{-4}$ & $9.8 \times 10^{-25}$ & $3.0 \times 10^{-26}$ \\
        $10^5$ & $10^6$ & $2.3 \times 10^{-6}$ & $1.6 \times 10^{-6}$ & $9.5 \times 10^{-27}$ & $2.9 \times 10^{-28}$ \\
        $10^6$ & $10^6$ & $2.2 \times 10^{-8}$ & $1.6 \times 10^{-8}$ & $9.3 \times 10^{-29}$ & $2.8 \times 10^{-30}$ \\ \hline
        \multicolumn{6}{c}{AE Aquarii} \\ \hline
        $10^5$ & $5 \times 10^7$ & $5.7 \times 10^{-3}$ & $4.0 \times 10^{-3}$ & $5.3 \times 10^{-23}$ & $1.6 \times 10^{-24}$ \\
        $10^6$ & $5 \times 10^7$ & $5.5 \times 10^{-5}$ & $3.9 \times 10^{-5}$ & $5.2 \times 10^{-25}$ & $1.6 \times 10^{-26}$ \\
        $10^7$ & $5 \times 10^7$ & $5.3 \times 10^{-7}$ & $3.8 \times 10^{-7}$ & $5.0 \times 10^{-27}$ & $1.5 \times 10^{-28}$ \\ \hline
        \multicolumn{6}{c}{V1460 Herculis} \\ \hline
        $10^4$ & $10^7$ & $2.3 \times 10^{-2}$ & $1.7 \times 10^{-2}$ & $5.5 \times 10^{-23}$ & $1.7 \times 10^{-24}$ \\
        $10^5$ & $10^7$ & $2.3 \times 10^{-4}$ & $1.6 \times 10^{-4}$ & $5.4 \times 10^{-25}$ & $1.6 \times 10^{-26}$ \\
        $10^6$ & $10^7$ & $2.2 \times 10^{-6}$ & $1.6 \times 10^{-6}$ & $5.2 \times 10^{-27}$ & $1.6 \times 10^{-28}$ \\
        $10^7$ & $10^7$ & $2.1 \times 10^{-8}$ & $1.5 \times 10^{-8}$ & $5.1 \times 10^{-29}$ & $1.5 \times 10^{-30}$ \\ \hline\hline
    \end{tabular}
\end{table*}

As shown in Table~\ref{tab:Tab_source_ampl}, if these WDs host accreted mountains with higher temperatures ($T \sim 10^6 - 10^7$~K), the maximum sustainable mass restricts the ellipticities to the range of $10^{-8} \lesssim \varepsilon_{\rm max} \lesssim 10^{-6}$. Consequently, the generated GW amplitudes fall below the detection thresholds of the considered instruments. One exception is AE Aquarii; at $T = 10^6$~K, this system can support an ellipticity of $\varepsilon_{\rm max} \approx 4 \times 10^{-5}$, producing a GW signal that remains detectable provided the WD is a nearly orthogonal rotator ($\alpha \sim 90^\circ$).

Figure~\ref{fig:h_freq_sources} displays the GW amplitudes for each system at two representative inclination angles ($\alpha = 10^\circ$ and $\alpha = 90^\circ$) alongside the sensitivity curves of space-based interferometers. The vertical bars span the adopted temperature range, with the upper and lower bounds corresponding to $T_{\rm min}$ and $T_{\rm max}$, respectively, as listed in Table~\ref{tab:Tab_source_ampl}. We find that the systems LAMOST~J0240+1952 and V1460~Her fall within the observable range across both inclination angles if their accreted mountains are relatively cool ($T \sim 10^4 - 10^5$~K). For CTCV~J2056-3014, the signal drops below detectability limits for a nearly aligned configuration ($\alpha \sim 10^\circ$). However, for $\alpha \sim 90^\circ$, this WD becomes observable provided the mountain is cold ($T = 10^4$~K), since a slightly hotter material ($T = 10^5$~K) yields an amplitude below the instrumental thresholds. Detection of AE Aquarii is feasible for mountain temperatures of $T \sim (10^5 - 10^6)$~K, assuming $\alpha \sim 90^\circ$.

In summary, the successful observation of these known sources by forthcoming space-based GW detectors strongly favors scenarios where the accreted matter is maintained at low to intermediate temperatures. We emphasize, however, that the resulting ellipticities, and thus the expected GW amplitudes, are highly sensitive to the polar magnetic field strengths of these objects, which in several cases remain observationally uncertain.

\section{Conclusion}
\label{sec:conclusions}
In this work, we investigated the emission of continuous gravitational waves by fast-spinning magnetic white dwarfs. We considered non-axisymmetric deformations arising from magnetically funneled accretion, using the model of magnetically confined mountains on WDs developed in detail by \citeauthor{rossetto2026magnetically} (\citeyear{rossetto2026magnetically}) \cite{rossetto2026magnetically} and summarized in Section \ref{sec:mountains_theory}. This model solves the magnetohydrodynamic (MHD) equilibrium equation --- the Grad-Shafranov equation \eqref{iso_grad_shafranov} --- to compute the mountain's structural properties, including its density, pressure, and magnetic field.

The gravitational wave amplitude emitted by a non-axisymmetric rotor depends on its mass, radius, rotation frequency, ellipticity, and distance from the observer (equation \eqref{AmpAC2}) \cite{maggiore2008gravitational}. Among these parameters, the ellipticity $\varepsilon$ is the quantity directly obtained from our MHD simulations. We found that $\varepsilon$ scales nearly linearly with the accreted mass $M_{\rm a}$ (Figure \ref{fig:ellip_per_Ma}), up to a ceiling set by the maximum mass that can be confined for a given stellar magnetic field $B_*$. At this maximum accreted mass $M_{\rm a, max}$, the star reaches a corresponding maximum ellipticity $\varepsilon_{\rm max}$. We find that $\varepsilon_{\rm max}$ scales directly with ${B_*}^2$ and inversely with the square of the mountain temperature $T$ (equation \eqref{max_mass_formula_fit}). For highly magnetized WDs ($B_* \sim 5\times10^{7}\,\mathrm{G}$) with relatively cool mountains ($T \sim 10^5\,\mathrm{K}$), ellipticities as large as $\sim 10^{-3}$ are achievable. These values represent upper limits, contingent on the WD having had sufficient time to accrete up to $M_{\rm a, max}$ from its companion.

Building upon these limiting ellipticities, we evaluated the corresponding upper limits of the continuous GW amplitude, $h_{\rm 0, max}$. Because the GW amplitude is directly proportional to the ellipticity, the numerical data allowed us to derive a parameterized scaling relation (Equation \ref{h_max_fit}), demonstrating that $h_{\rm 0, max}$ mirrors the functional dependence of $\varepsilon_{\rm max}$, scaling directly with $B_*^2$ and inversely with $T^2$. This relation can be generalized to explicitly incorporate the dependencies on the GW frequency, $f$, and the source distance, $d$, as follows:
\begin{align}
    &h_{\rm 0, max} = (3.0 \pm 0.2) \times 10^{-23} \left(\frac{T}{10^5 \,\mathrm{K}}\right)^{-2} \left(\frac{B_*}{5\times 10^7 \,\mathrm{G}}\right)^{2} \nonumber \label{h_fit_gen}
    \\
    & \left(\frac{f}{0.03 \,\mathrm{Hz}}\right)^2 \left(\frac{100 \,\mathrm{pc}}{d}\right). 
\end{align}
The predicted GW amplitudes span a broad range, reaching up to $\sim 10^{-21}$ for the most massive sustainable mountains, placing these optimal configurations well within the sensitivity bands of planned space-based interferometers.

By exploring a comprehensive parameter space, including rotation periods ($10 \,\mathrm{s} \leq P \leq 100 \,\mathrm{s}$), inclination angles ($\alpha$), and source distances ($10 \, \mathrm{pc} \leq d \leq 1 \, \mathrm{kpc}$), we mapped the detectability horizons for future GW missions. We found that LISA, Taiji, and TianQin are primarily sensitive to highly deformed systems ($\varepsilon_{\rm max} \gtrsim 10^{-4}$), typically requiring fast rotation ($P \lesssim 30$~s) and nearly orthogonal alignments ($\alpha \approx 90^\circ$). Conversely, the proposed next-generation observatories BBO and DECIGO present significantly higher sensitivities, capable of measuring signals from WDs with ellipticities down to $\varepsilon \sim 10^{-6}$, provided the star spins sufficiently fast. In all cases, an important parameter degeneracy underlies these prospects: achieving the macroscopic states required for wider detectability depends on specific $(T, B_*)$ combinations that preserve the magnetohydrodynamic stability of the accreted matter.

To ground our theoretical framework in observational prospects, we applied this modeling to four known magnetic cataclysmic variables hosting fast-spinning WDs: LAMOST~J0240+1952, V1460~Her, CTCV~J2056-3014, and AE~Aquarii. Our analysis demonstrates that the detectability of these actual astrophysical systems is fundamentally governed by the thermal state of the accreted plasma. While LAMOST~J0240+1952 and V1460~Her are observable across various inclination angles if the mountain is relatively cool ($T \sim 10^4 - 10^5$~K), CTCV~J2056-3014 and AE~Aquarii demand more restrictive, nearly orthogonal configurations ($\alpha \approx 90^\circ$). Overall, the most viable candidates for future space-borne missions are those harboring cooler accreted mountains. Nonetheless, precise GW forecasts for individual objects will ultimately require more robust observational constraints, as the amplitude estimations remain dependent on surface magnetic fields that are currently not well established.

The mountain model employed here admits several natural extensions toward more realistic physics. We adopted an isothermal equation of state for simplicity, but future work should explore polytropic equations of state to capture more physically motivated regimes, such as the degenerate electron gas \cite{priymak2011quadrupole,zhang2009there}. A stability analysis of the equilibrium solutions is also needed to assess their long-term dynamical viability. This could be carried out using full MHD solvers such as \texttt{PLUTO} or \texttt{ATHENA++}, with the Grad-Shafranov equilibria serving as initial conditions \cite{mignone2012pluto, stone2020athena}. Such simulations would simultaneously allow investigation of plasma instabilities of Parker, interchange, and ballooning type \cite{payne2007burial,mukherjee2013mhda,mukherjee2013mhdb}, additionally, instability analyses could be carried out with semi-analytical models \cite{kulsrud2020anomalous,brunet2026relaxation}. Finally, the mountain equilibria were computed here within Newtonian gravity. We derived the general relativistic correction in Appendix \ref{app:compactness} and confirmed that the Newtonian treatment is adequate for the white dwarf scenario.

\section*{Acknowledgments}
This study was financed, in part, by the São Paulo Research Foundation (FAPESP), Brasil. Process Number 2024/21854-2. MFS is also grateful for the financial support of FAPESP (grants No. 2026/06733-0 and No. 2021/01089-1). DFG also thanks FAPESP for grant 2021/02120-0, and Conselho Nacional de Desenvolvimento Científico e Tecnológico (CNPq) for grant 304574/2024-4. The computations were performed using the Hydra HPC Cluster at the Laboratory of Theoretical Astrophysics, University of São Paulo, supported by FAPESP grant 2022/03972-2. We thank Ryan Brunet, Prof. Grzegorz Kowal and Prof. Andrew Melatos for valuable discussions about this work.

\printbibliography

\appendix

\section{Compactness of White Dwarfs and General Relativistic Corrections}
\label{app:compactness}
The equilibrium solutions that we encounter in our simulations depend, in a non-trivial way, on the mass $M_*$ and radius $R_*$ of the white dwarf. The combination $M_*/{R_*}^2$ is present in the gravitational field $\Phi$, which is present in the main equations of the problem -- that is, equations \eqref{iso_grad_shafranov} and \eqref{Fpsi}. Furthermore, the combination $M_*{R_*}^2$ is encountered in the pre-accretion moment of inertia $I_0$, and it is in the definition of the ellipticity $\varepsilon$, equation \eqref{epsilon}. Lastly, the ratio $M_*/R_*$ is present in the compactness $\mathcal{C}$, defined by
\begin{equation}
    \label{compactness}
    \mathcal{C} = \frac{GM_*}{c^2R_*},
\end{equation}
and it is the term that quantifies how ``general relativistic'' the system is. Thus, in this appendix we investigate the role of the star's mass-radius relation on  $\varepsilon$ and on the general relativistic corrections.

\begin{figure}
    \centering
    \includegraphics[width=0.49
    \textwidth]{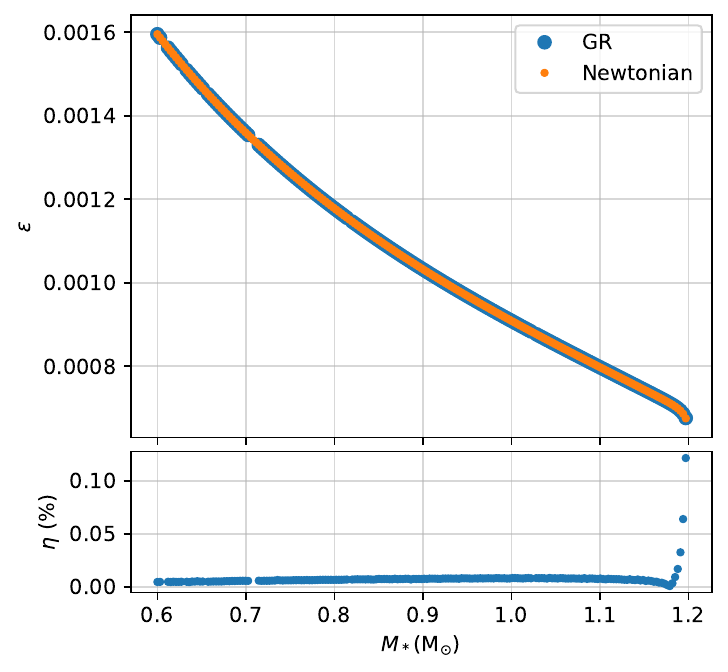}
    \caption{Ellipticity $\varepsilon$ as a function of the stellar mass $M_*$, following the mass-radius relation \eqref{mass_radius}. For all simulations, $M_{\rm a} = 10^{-3} M_\odot$, $B_* = 5\times10^7\,\mathrm{G}$, $b=10$ and $T = 1.59\times10^5\,\mathrm{K}$. Blue and orange circles represent the general relativistic and Newtonian solutions, respectively. The bottom plot shows the fractional difference $\eta$ in ellipticity between the two solutions.}
    \label{fig:ellip_per_stellar_mass}
\end{figure}

White dwarfs have an equation of state much better understood than neutron stars. A successful analytical approximation for the radius-mass relation of white dwarfs is
\begin{equation}
    \label{mass_radius}
    R_*(M_*) = a \left(\frac{M_*}{b}\right)^{-1/3}\left[ 1 - \left(\frac{M_*}{b}\right)^{4/3}\right]^{1/2},
\end{equation}
where $a = 7.8\times 10^8\,\mathrm{cm}$ and $b = 1.44 M_\odot$ \cite{nauenberg1972analytic,dufour2017montreal}. Therefore, we use this expression to obtain a consistent value of $R_*$ given a particular value of $M_*$. On the main plot of Figure \ref{fig:ellip_per_stellar_mass}, we show the behavior of the ellipticity $\varepsilon$ with the stellar mass $M_*$, and its corresponding $R_*$ value, for $M_{\rm a} = 10^{-3}M_\odot$, $B_* = 5\times10^7\,\mathrm{G}$ and $T = 1.59\times10^5\,\mathrm{K}$. We notice from the figure that the ellipticity decreases with increasing mass, approximately halving its value from $0.6M_\odot$ to $1.1M_\odot$.  A theoretical prediction or explanation of this behavior of the ellipticity is challenging given the multi-factorial way that mass-radius relationship affects the system, as explained previously. 

Using the formalism in \citeauthor{rossetto2023magnetically} \citeyear{rossetto2023magnetically}, we can compute the Grad-Shafranov equilibrium in general relativity \cite{rossetto2023magnetically}. On the main plot of Figure \ref{fig:ellip_per_stellar_mass}, we show the general relativistic results with big blue circles and the Newtonian ones with small orange circles. The curves are nearly identical. To display the quantitative difference between both results, on the bottom panel of Figure \ref{fig:ellip_per_stellar_mass}, we plot the fractional difference $\eta$ in ellipticity, i.e. $\eta = |\varepsilon_{\rm GR} - \varepsilon_{\rm N}|/\varepsilon_{\rm GR}$ with $\varepsilon_{\rm GR}$ being the relativistic result and $\varepsilon_{\rm N}$ being the Newtonian result. For most of the mass range ($\lesssim 1.17 M_\odot$), the fractional difference is less than $0.02\%$. Nevertheless, as the mass approaches $1.2M_\odot$, the difference increases rapidly but stays under $\sim0.13\%$. Therefore, even for highly compact white dwarfs, the general relativistic corrections are quite small, justifying the Newtonian approach that was assumed throughout this paper. For neutron stars, the percentage difference in ellipticity can be as high as $20\%$, showing that while relativistic corrections can be safely neglected for white dwarfs, they are fundamental for neutron stars \cite{rossetto2025quadrupole}. This difference is expected since the compactness of white dwarfs is $\mathcal{C}\sim2\times10^{-4}$ and for neutron stars $\mathcal{C}\sim 2\times10^{-1}$.

\end{document}